\documentclass{iaus}
\usepackage{graphicx}

\title[ELSA] 
{ELSA: An integrated, semi-automated nebular abundance package}

\author[Johnson et al.]   
{Matthew D. Johnson$^1$,
 Jesse S. Levitt$^2$, \break Richard B. C. Henry$^3$, \and Karen B. Kwitter$^2$}

\affiliation{$^1$Department of Astronomy, Wesleyan University,
Middletown, CT 06459, USA\break email: mdjohnson@wesleyan.edu \\[\affilskip]
$^2$Department of Astronomy, Williams College, Williamstown, MA 01267, USA\break email: 08jsl@williams.edu, kkwitter@williams.edu \\[\affilskip] $^3$H. L. Dodge Department of Physics and Astronomy, U. Oklahoma, Norman, OK 73019, USA\break email: henry@nhn.ou.edu}

\pubyear{2006}
\volume{234}  
\pagerange{119--126}
\date{?? and in revised form ??}
\jname{Planetary Nebulae in Our Galaxy and Beyond}
\editors{M.J. Barlow \& R.H. M\'endez, eds.}
\begin{document}

\maketitle


\section{Features}

We present ELSA, a new modular software package, written in C, to analyze and manage spectroscopic data from emission-line objects. In addition to calculating plasma diagnostics and abundances from nebular emission lines, the software provides a number of convenient features including the ability to ingest logs produced by IRAF's {\it splot} task, to semi-automatically merge spectra in different wavelength ranges, and to automatically generate various data tables in machine-readable or \LaTeX~format. ELSA features a highly sophisticated interstellar reddening correction scheme that takes into account temperature and density effects as well as He II contamination of the hydrogen Balmer lines. Abundance calculations are performed using a 5-level atom approximation with recent atomic data, based on R. Henry's \textsc{abun} program. Downloading and detailed documentation for all aspects of ELSA are available at http://www.williams.edu/Astronomy/research/PN/.

\subsection{Input}
ELSA as a tool for astrophysicists has a two-fold purpose: it automates some of the more laborious aspects of spectral data analysis for emission-line objects, while also providing highly accurate and refined facilities for dereddening corrections and atomic abundance calculations. Furthermore, it can generate output to screen, to spreadsheet delimited file formats, or to \LaTeX~source files for each step of the analysis process.

ELSA takes the logfile output from IRAF's {\it splot} task and matches the measured wavelengths (adjustable for a user-input radial velocity) to a list of canonical reference wavelengths provided by the user. ELSA can then perform a variety of tasks, including interstellar reddening correction, generation of a wavelength-flux data table, computation of plasma diagnostics and elemental abundances, and flagging unknown or commented wavelengths for later review. User-defined comments in the original logfiles are preserved throughout the analysis process. It is also possible to bypass the reddening correction phase in order to run analysis of data that has been previously dereddened independent of ELSA.

\subsection{Reddening Correction \& T$_e$-N$_e$ Loops}

Dereddening is performed through an iterative process that works in several stages: correcting for the He$^{+2}$ Pickering line contamination of the H$\mathrm{\alpha}$ and H$\mathrm{\beta}$ lines, accounting for the co-dependence of  T$_e$ and N$_e$ determinations, dynamically calculating the appropriate H$\mathrm{\alpha}/$H$\mathrm{\beta}$ ratio  based on T$_e$ and N$_e$, and calculating the reddening coefficient $c$ based on the appropriate intrinsic Balmer ratio. ELSA makes a first-pass correction assuming a user-defined default H$\mathrm{\alpha}/$H$\mathrm{\beta}$ ({\it e.g.}, 2.86). This rough value is used to apply a reddening correction from \cite[Savage \& Mathis (1979)]{Savage79} to H$\mathrm{\alpha}$ and to He II $\mathrm{\lambda}4686$. The corrected $\mathrm{\lambda}4686$ flux is then used to calculate and then subtract the portion of the fluxes at H$\mathrm{\alpha}$ and  H$\mathrm{\beta}$ that come from He$^{+2}$ Pickering line contamination. The reddening correction is then run inversely to ``reredden''  H$\mathrm{\alpha}$ and  H$\mathrm{\beta}$. A new value of $c$ is then computed and the process repeats until  successive values of $c$ converge. The final $c$ value is used to deredden [O III] $\mathrm{\lambda}4363$ and $\mathrm{\lambda}5007$, which are used to derive T$_e$ assuming a fixed ``seed'' N$_e$. A 2D polynomial interpolation table is used to determine N$_e$ using the previously calculated T$_e$. A new  T$_e$ is then calculated based on the improved N$_e$, and so on until both approach stable values.

The final values for T$_e$ and N$_e$ are used to refine the intrinsic H$\mathrm{\alpha}/$H$\mathrm{\beta}$ ratio, using data and code based on \cite[Storey \& Hummer (1995)]{storey95}. ELSA then returns to the beginning of the loop and makes a second pass at dereddening the raw data with the new H$\mathrm{\alpha}/$H$\mathrm{\beta}$ used to determine $c$. The entire loop continues to iterate in this manner until H$\mathrm{\alpha}/$H$\mathrm{\beta}$ itself converges satisfactorily. The final $c$ value is then used to deredden \textit{all} lines read from the raw data. ELSA then stores this as the pool of data to be used for generating tables or computing chemical abundances. 

\subsection{Chemical Abundance Calculations}

ELSA uses a standard five-level atom model as described in \cite[Kwitter \& Henry (2001)]{kwitter01}, with code based on R. B. C. Henry's \textsc{abun} routine. ELSA can compute abundances for various ionization stages of He, N, O, S, Ar and Ne; total abundances are computed using ionization correction factors as described in Kwitter \& Henry (2001). Atomic data are drawn principally from \cite[Osterbrock \& Ferland (2005)]{osterbrock05} or from \cite[Mendoza (1983)]{mendoza83}. Currently, ELSA contains a two-zone ionization model, with T$_{[N~II]}$ and T$_{[O~III]}$ regions. 

\subsection{Tabular Output Options}

ELSA can produce output from its tasks in variety of formats. The dereddened line fluxes can be output to a tab-delimited file, a comma-separated value (CSV) file readable by spreadsheet programs such as Microsoft Excel, or a \LaTeX~source file suitable for insertion into a publication manuscript. The content of this table is controlled by a user-defined list of wavelengths, and corrected line intensities are given relative to H$\mathrm{\beta}$=100. ELSA also has the capability to produce a composite table of the same lines across multiple objects in \LaTeX~format, a product that would otherwise be quite laborious to produce. Comments read from log files are also preserved and output into the tabular results

\subsection{Planned Improvements}

Future plans for ELSA include the incorporation of robust error propagation, the addition of UV and IR analyses, expansion to a multiple-zone model (similar to IRAF's {\it nebular} package) and support to accommodate extragalactic emission-line objects.

\begin{acknowledgments}
We are grateful to the Keck Northeast Astronomy Consortium, which is partially supported by the National Science Foundation, and to the Bronfman Science Center and Astronomy Department of Williams College.  We also acknowledge National Science Foundation grant AST 03-07118 to the University of Oklahoma.
\end{acknowledgments}

\end{document}